\documentclass[a4paper]{article}
\usepackage{graphicx}
\usepackage{amsmath}

\begin{document}

\newcommand{\proba}{\ensuremath{\mathcal{P}}}
\newcommand{\eline}{E_{\gamma J,\gamma'J'}}
\newcommand{\gj}{\gamma J}
\newcommand{\gjp}{\gamma' J'}

\huge

\begin{center}
The hybrid opacity code SCO-RCG: recent developments
\end{center}

\vspace{0.5cm}

\large

\begin{center}
Jean-Christophe Pain\footnote{jean-christophe.pain@cea.fr (corresponding author)}, Franck Gilleron$^1$, Quentin Porcherot$^2$ and Thomas Blenski$^3$
\end{center}

\vspace{0.2cm}

\normalsize

\begin{center}
$^1$CEA, DAM, DIF, F-91297 Arpajon, France

$^2$DGA, 94110 Arcueil, France

$^3$CEA, DSM, IRAMIS, F-91191 Gif-sur-Yvette, France
\end{center}

\vspace{0.5cm}

The proper accounting for spectral lines in hot, partially ionized matter is important for opacities and plasma diagnostics. The number of levels and electric-dipolar (E1) lines is huge, especially for medium or high-$Z$ elements in the case of electron configurations with one or more open subshells. Fortunately, lines usually coalesce into broad unresolved patterns due to their physical broadening mechanisms (radiative decay, Doppler, Stark, auto-ionization, ...). Theoretical estimate for the statistical width of a transition array (TA) was given by Moszkowski \cite{Moszkowski62} and in a series of papers by Bauche \emph{et al.} (see \cite{Bauche88}). In such an approach, each TA can be modelled statistically by a continuous envelope whose first two moments (centre of gravity and variance) are calculated using the second-quantization techniques of Judd, Racah algebra and Jucys' graphical methods. Resulting patterns are named unresolved transition arrays (UTAs) in intermediate coupling and spin-orbit split arrays (SOSAs) in relativistic $jj$ coupling. Spectra obtained by such global methods are made of unresolved structures which are not sufficiently detailed in some cases. At low density, many TAs are not ``smooth'', but exhibit some variations (signature of lines). The Rosseland mean is very sensitive to the resulting hollows in the spectrum, whose porosity (localized absence of lines) makes the plasma transparent at several frequencies. Lines can subsequently play an important role in the modelling of radiation transport. However, their computation requires complex numerical calculations, based on a detailed description of atomic structure, implying the diagonalization of the Hamiltonian matrix. 

We therefore decided to develop the hybrid SCO-RCG code, assuming local thermodynamic equilibrium and representing the best compromise between precision and calculation time \cite{Porcherot11}. The statistical part is calculated by SCO (Super-Configuration Opacity) code \cite{Blenski00}, relying on the concept of ``super-configuration'' (SC), for instance $\left[1s2s2p\right]^3\left[3s\right]^1\left[3p3d4s\right]^4$. In this way, a rather limited number of SCs can represent a huge number of ordinary configurations. The ensemble of lines associated to a mono-electronic jump between two SCs represents a super-transition array (STA \cite{Barshalom89}) modelled by a statistical distribution whose first two moments can be expressed in terms of partition functions, evaluated by recursive techniques. In order to decide whether a detailed treatment of lines is necessary and to determine the validity of statistical methods, the code uses criteria involving characteristics of the distribution of lines (number, moments, individual widths, energy-amplitude correlation). Data required for the calculation of detailed TAs (Slater, spin-orbit and dipolar integrals) are calculated by SCO, which takes into account plasma screening and density effects on wavefunctions (proper treatment of pressure ionization). These data are used to compute level energies and lines by RCG routine of Cowan's code \cite{Cowan81}, modified for our purposes.  The TAs for which a detailed treatment is not required or impossible are described statistically, by UTA, SOSA or STA formalisms. In SCO-RCG, configuration mixing is limited to electrostatic interaction between relativistic sub-configurations ($n\ell j$ orbitals) belonging to a non-relativistic configuration ($n\ell$ orbitals). SCO-RCG calculations are restricted to a particular type of SC, in which all supershells are made of individual orbitals up to a limit beyond which all the remaining orbitals are gathered into a large final supershell $\sigma_M$, for instance

\begin{equation}
\sigma_1=\left[1s\right]^{N_1}, \sigma_2=\left[2s\right]^{N_2}, \sigma_3=\left[2p\right]^{N_3},\cdots, \sigma_{M-1}=\left[8s\right]^{N_{M-1}}, \sigma_M=\left[\left(8p\right)\cdots\cdots k_{\mathrm{max}}\right]^{N_M}.
\end{equation}

If $N_M=0$, the calculation involves only detailed configurations and therefore consists only of DLAs, UTAs or SOSAs, depending on whether TAs can be detailed or not. If $N_M\ge 1$, the spectra consists only of STAs. The first orbital $k_1$ of $\sigma_M$ ($8p$ in example (1)) is determined consistently with Inglis-Teller limit \cite{Inglis39} and in order to minimize the contribution of $\sigma_M$ to the partition function. In order to complement DLA efforts, the code was recently improved with the PRTA (Partially Resolved Transition Array) model \cite{Iglesias12}, which may replace the single feature of a UTA by a small-scale detailed TA that conserves the known TA properties (energy and variance) and yields improved high-order moments (skewness, kurtosis, ...). In the PRTA approach, open subshells are split in two groups. The main group includes the active electrons and those electrons that couple strongly with the active ones. The other subshells are relegated to the secondary group. A small scale DLA calculation is performed for the main group (assuming therefore that the subshells in the secondary group are closed) and a statistical approach for the secondary group assigns the missing UTA variance to the lines.

The total opacity is the sum of photo-ionization, inverse Bremsstrahlung and Thomson scattering spectra calculated by SCO code and a photo-excitation spectrum arising from contributions of SCO and Cowan's codes (see Fig. 1) in the form

\begin{equation}
\kappa(h\nu)=\frac{1}{4\pi\epsilon_0}\frac{\mathcal{N}}{A}\frac{\pi e^2h}{mc}\sum_{X\rightarrow X'}f_{X\rightarrow X'}\mathcal{P}_X\Psi_{X\rightarrow X'}(h\nu),
\end{equation}

\noindent where $h$ is Planck's constant, $\mathcal{N}$ the Avogadro number, $\epsilon_0$ the vacuum polarizability, $m$ the electron mass, $A$ the atomic number and $c$ the speed of light. $\mathcal{P}$ is a probability, $f$ an oscillator strength, $\Psi(h\nu)$ a profile and the sum $X\rightarrow X'$ runs over lines, UTAs or STAs. Special care is taken to calculate the probability of the corresponding ``object'' $X$ (level $\gamma J$, configuration $C$ or SC $\Xi$) because it can be the starting point for different transitions (DLA, UTA, ...). In order to ensure the normalization of probabilities, configurations are first split into three disjoint ensembles: $\mathcal{D}$ (configurations whose levels can be calculated), $\mathcal{C}$ (configurations too complex to be detailed) and $\mathcal{S}$ (configurations gathered into SCs). The total partition function then reads

\begin{equation}
U_{\mathrm{tot}}=U\left(\mathcal{D}\right)+U\left(\mathcal{C}\right)+U\left(\mathcal{S}\right),
\end{equation}

\noindent where each term is a trace over quantum states of the form Tr$\left[e^{-\beta\left(\hat{H}-\mu\hat{N}\right)}\right]$ in the corresponding ensemble ($\hat{H}$ is the Hamiltonian, $\hat{N}$ the number operator, $\mu$ the chemical potential and $\beta=1/\left(k_BT\right)$). The probability $\mathcal{P}_X$ of species $X$ reads
\begin{equation}
\begin{array}{l}
\mathcal{P}_C=\left\{\begin{array}{l}
\frac{1}{U_{\mathrm{tot}}}g_C~e^{-\beta\left(E_C-\mu N_C\right)}\;\;\;\mathrm{if}\;\;\;C\in\mathcal{C}\\
\frac{1}{U_{\mathrm{tot}}}\sum_{\gamma J\in C}\left(2J+1\right)e^{-\beta\left(E_{\gamma J}-\mu N_C\right)}\;\;\;\mathrm{if}\;\;\;C\in\mathcal{D},\\
\end{array}\right.\\
\mathcal{P}_{\gamma J}=\frac{1}{U_{\mathrm{tot}}}\left(2J+1\right)e^{-\beta\left(E_{\gamma J}-\mu N_C\right)}\;\;\;\mathrm{if}\;\;\;\gamma J\in C\in\mathcal{D},\\
\mathcal{P}_{\Xi}=\frac{1}{U_{\mathrm{tot}}}\sum_{C\in\Xi}g_C~e^{-\beta\left(E_C-\mu N_C\right)}\;\;\;\mathrm{if}\;\;\;C\in\Xi\in\mathcal{S},
\end{array}
\end{equation}

\noindent so that

\begin{equation}
\sum_{\gamma J\in\mathcal{D}}\mathcal{P}_{\gamma J}+\sum_{C\in\mathcal{C}}\mathcal{P}_C+\sum_{\Xi\in\mathcal{S}}\mathcal{P}_{\Xi}=1.
\end{equation}

In the case where the transition $C\rightarrow C'$ is a UTA that can be replaced by a PRTA (see Fig. 2), its contribution to the opacity is modified according to

\begin{equation}
f_{C\rightarrow C'}~\mathcal{P}_C~\Psi_{C\rightarrow C'}(h\nu)\approx\sum_{\bar{\gamma}\bar{J}\rightarrow\bar{\gamma'}\bar{J'}}f_{\bar{\gamma}\bar{J}\rightarrow\bar{\gamma'}\bar{J'}}~\mathcal{P}_{\bar{\gamma}\bar{J}}~\Psi_{\bar{\gamma}\bar{J}\rightarrow\bar{\gamma'}\bar{J'}}(h\nu),
\end{equation}

\noindent where the sum runs over PRTA lines $\bar{\gamma}\bar{J}\rightarrow\bar{\gamma'}\bar{J'}$ between all ``virtual'' levels of the reduced set of orbitals, $f_{\bar{\gamma}\bar{J}\rightarrow\bar{\gamma'}\bar{J'}}$ is the corresponding oscillator strength and $\Psi_{\bar{\gamma}\bar{J}\rightarrow\bar{\gamma'}\bar{J'}}$ is the line profile augmented with the statistical width due to the other (non included) spectator subshells. The probability of a virtual level $\bar{\gamma}\bar{J}$ belonging to the fake configuration $\bar{C}$ reads

\begin{equation}
\mathcal{P}_{\bar{\gamma}\bar{J}}=\frac{\left(2\bar{J}+1\right)e^{-\beta\left(E_{\bar{\gamma}\bar{J}}-\mu N_C\right)}}{\sum_{\bar{\gamma}\bar{J}\in\bar{C}}\left(2\bar{J}+1\right)e^{-\beta \left(E_{\bar{\gamma}\bar{J}}-\mu N_C\right)}}\times\mathcal{P}_C\;\;\;\;\mathrm{with}\;\;\;\;\sum_{\bar{\gamma}\bar{J}\in\bar{C}}\mathcal{P}_{\bar{\gamma}\bar{J}}=\mathcal{P}_C,
\end{equation}

\noindent where $\mathcal{P}_C$ is given in Eq. (4).

The SCO-RCG code has already been used for astrophysical applications \cite{Gilles11} and interpretation of experimental spectra (see Fig. 3). The PRTA model, recently extended to the hybrid statistical / detailed approach, enables one to reduce the statistical part and fastens the calculations. We plan to improve the treatment of Stark broadening in order to increase the capability of the code as concerns K-shell spectroscopy.

\clearpage

\begin{figure}
\begin{center}
\includegraphics[width=12cm]{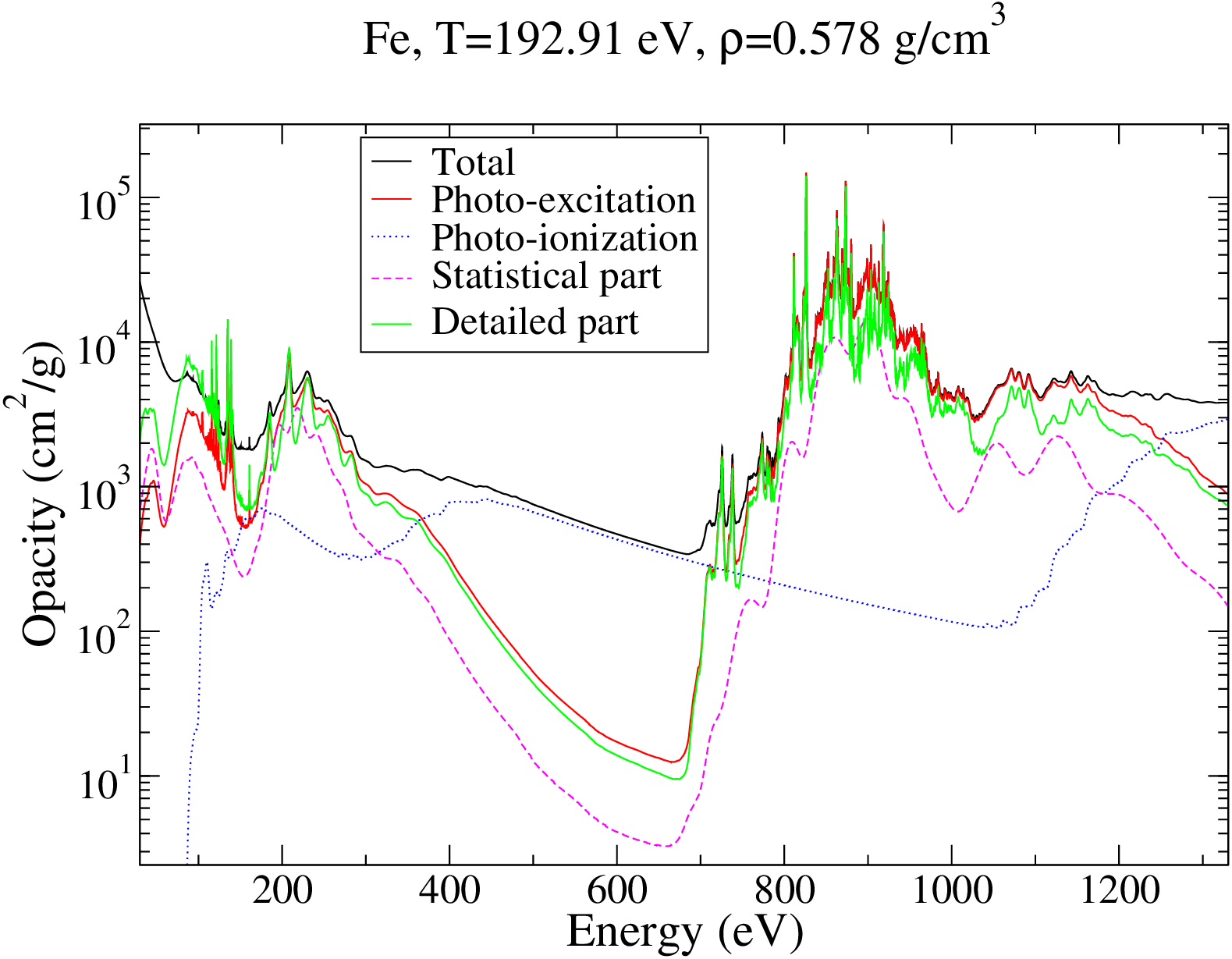}
\caption{Contributions to opacity in a SCO-RCG calculation \cite{Porcherot11} (boundary of the convective zone of the Sun).}
\label{fig1}
\end{center}
\end{figure}

\clearpage

\begin{figure}
\begin{center}
\includegraphics[width=12cm]{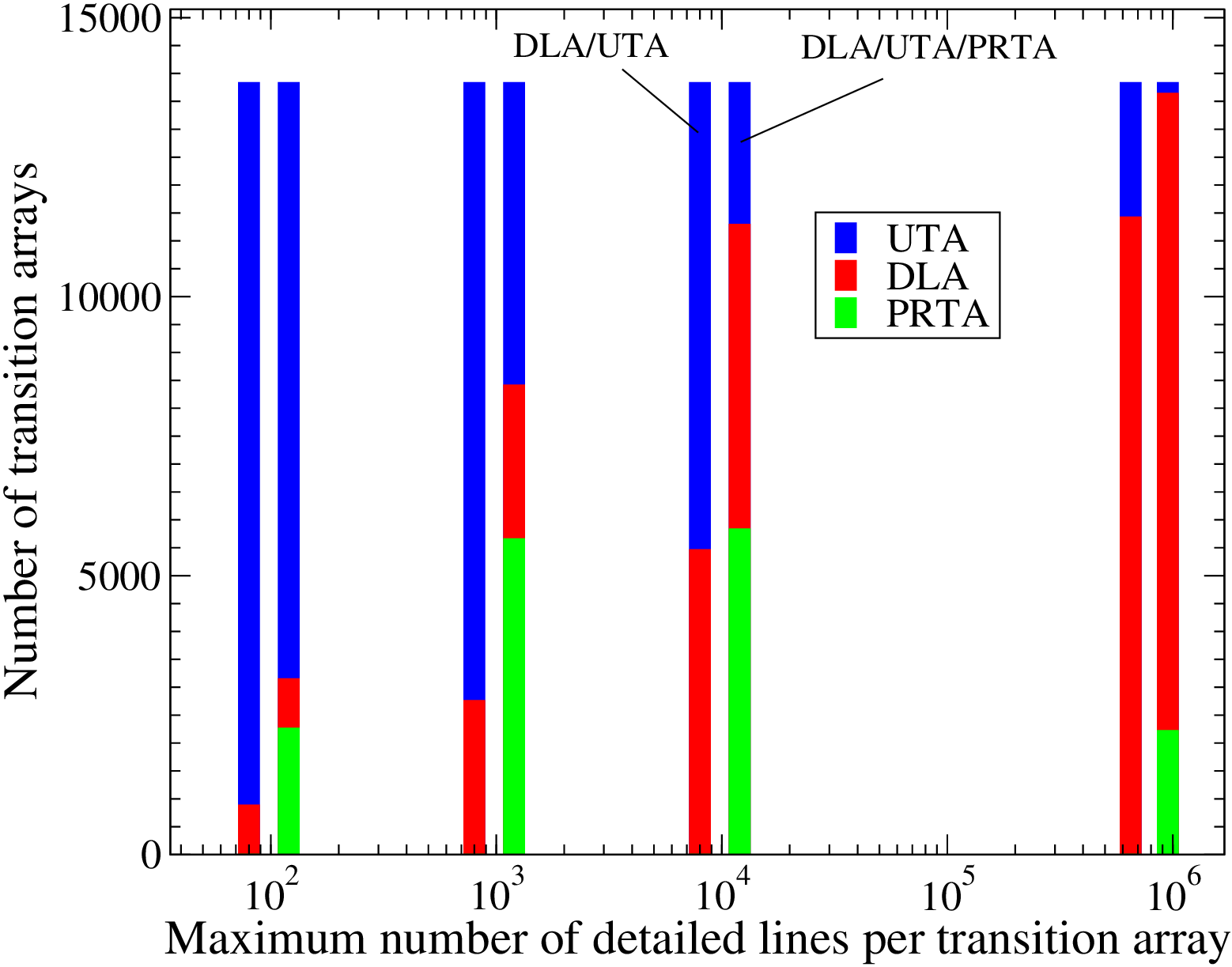}
\caption{Distribution of TAs (case of Fig. 1).}
\label{fig2}
\end{center}
\end{figure}

\clearpage

\begin{figure}
\begin {center}
\includegraphics[width=12cm]{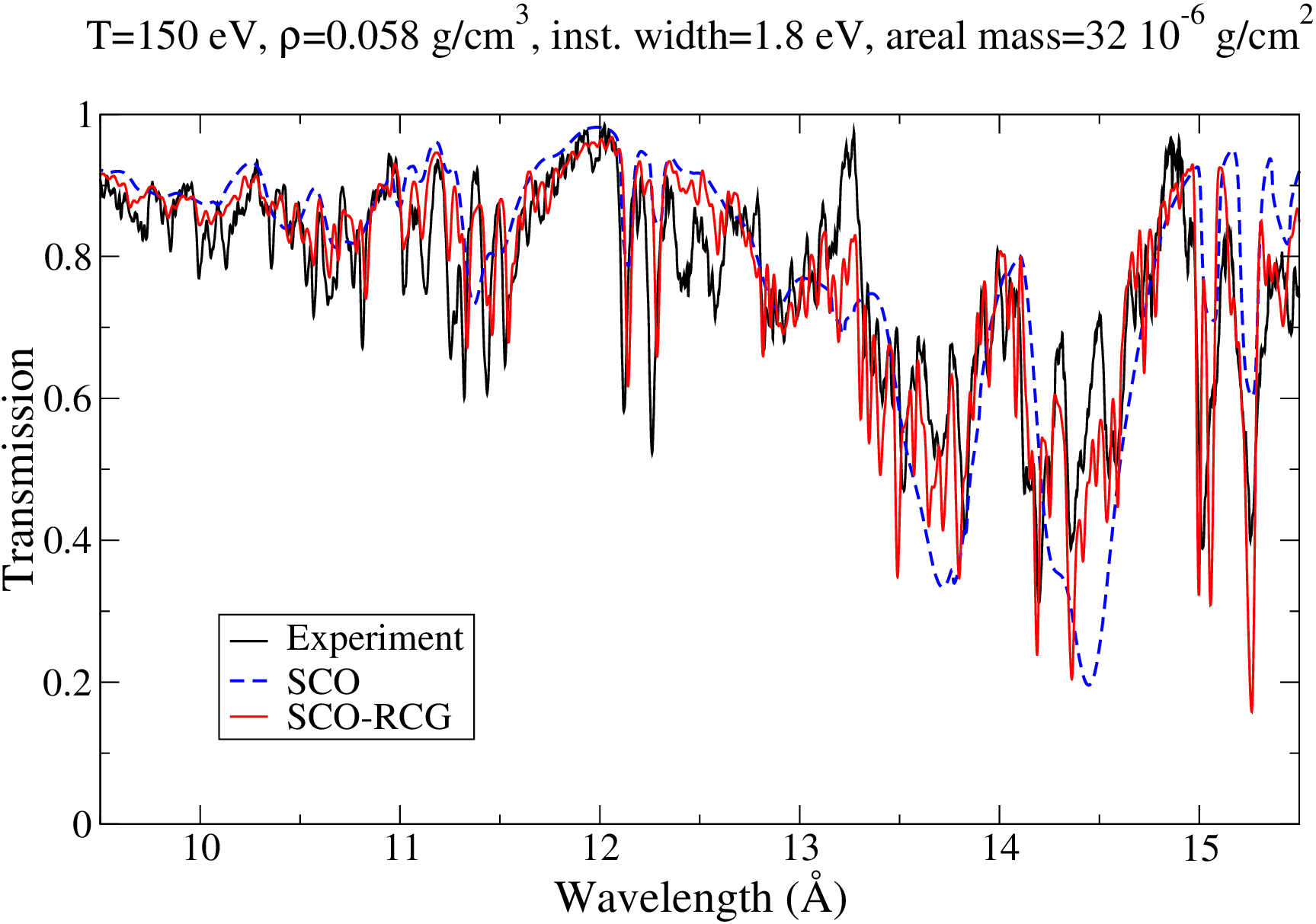}
\caption{Experimental spectrum obtained by Bailey \emph{et al.} \cite{Bailey07} compared to a SCO-RCG calculation \cite{Porcherot11}.}
\label{fig3}
\end{center}
\end{figure} 

\end{document}